%% file: IEEE_IoT_Journal.tex
\documentclass[journal]{IEEEtran}
\usepackage{amsmath,amsfonts, amssymb}
\usepackage{algorithmic}
\usepackage{array}
\usepackage{textcomp}
\usepackage{stfloats}
\usepackage{url}
\usepackage{verbatim}
\usepackage{graphicx}
\def\BibTeX{{\rm B\kern-.05em{\sc i\kern-.025em b}\kern-.08em
    T\kern-.1667em\lower.7ex\hbox{E}\kern-.125emX}}
\usepackage{balance}
\usepackage{subcaption}
\usepackage[table]{xcolor}
\usepackage{multirow}
\usepackage{enumitem}

\definecolor{sanoBlue}{HTML}{5F87FF}
\definecolor{sanoGreen}{HTML}{10AC84}
\definecolor{sanoLightRed}{HTML}{FF5F87}

\newcommand\ltnorm[1]{\lVert#1\rVert_{2}}
\newcommand\lossfn[1]{#1$_{\ell}$}

\begin{document}
\title{Proof of Reasoning for Privacy Enhanced Federated Blockchain Learning at the Edge}

\author{%
James Calo, Benny Lo \IEEEmembership{Senior Member, IEEE}
\thanks {James Calo (Department of Computing) and Benny Lo (Department of Surgery and Cancer) are with the Hamlyn Centre, Imperial College London, South Kensington, London, UK. Corresponding author (James Calo) e-mail: jam414@ic.ac.uk.}
\thanks {Copyright (c) 2026 IEEE. Personal use of this material is permitted. However, permission to use this material for any other purposes must be obtained from the IEEE by sending a request to pubs-permissions@ieee.org.}
}

\markboth{IEEE Internet of Things Journal,~Vol.~13, No.~11, January~2026}%
{James Calo, Benny Lo: Privacy Enhanced Blockchain Federated Learning at
the Edge}

\maketitle

\input{01_Abstract/01_abstract}

\input{02_Introduction/01_introduction}

\input{02_Introduction/02_system_overview}

\input{03_RelatedWork/01_related_work}

\input{04_Methods/01_methods} 

\input{05_Results/01_results} 

\input{06_Conclusion/01_conclusion} 

\bibliographystyle{IEEEtran}
\bibliography{IoTJ_2024}

\end{document}

%% file: 01_Abstract/01_abstract.tex
\begin{abstract}
Consensus mechanisms are the core of any blockchain system. However, the majority of these mechanisms do not target federated learning directly nor do they aid in the aggregation step. This paper introduces Proof of Reasoning (PoR), a novel consensus mechanism specifically designed for federated learning using blockchain, aimed at preserving data privacy, defending against malicious attacks, and enhancing the validation of participating networks. Unlike generic blockchain consensus mechanisms commonly found in the literature, PoR integrates three distinct processes tailored for federated learning. Firstly, a masked autoencoder (MAE) is trained to generate an encoder that functions as a feature map and obfuscates input data, rendering it resistant to human reconstruction and model inversion attacks. Secondly, a downstream classifier is trained at the edge, receiving input from the trained encoder. The downstream network's weights, a single encoded datapoint, the network's output and the ground truth are then added to a block for federated aggregation. Lastly, this data facilitates the aggregation of all participating networks, enabling more complex and verifiable aggregation methods than previously possible. This three-stage process results in more robust networks with significantly reduced computational complexity, maintaining high accuracy by training only the downstream classifier at the edge. PoR scales to large IoT networks with low latency and storage growth, and adapts to evolving data, regulations, and network conditions.
\end{abstract}


%% file: 02_Introduction/01_introduction.tex
\section{Introduction}

\IEEEPARstart{D}{eep} learning has emerged as the dominant paradigm for neural network development in recent years \cite{RN136, RN138}. However, training deep networks directly on IoT (Internet of Things) devices at the edge remains impractical due to their substantial computational requirements \cite{RN93, RN135}. Consequently, the vast and largely untapped computational potential of IoT devices, particularly in healthcare, is underutilized. These devices, ubiquitous in medical settings, represent an opportunity to decentralize artificial intelligence by shifting computation from centralized cloud systems to the edge, leveraging existing mist architectures already deployed in hospitals. This shift has the potential to revolutionize healthcare Artificial Intelligence (AI) by enabling real-time processing, reducing latency, and enhancing data security.
\par
Nevertheless, this transition is fraught with challenges. Even with access to powerful computational resources, deep learning models are prone to overfitting on large datasets comprising millions of images \cite{RN130}. Furthermore, successful training hinges on the availability of vast amounts of annotated data \cite{RN58}, much of which is inaccessible in healthcare due to stringent privacy regulations and data imbalances. These limitations underscore one of the most pressing challenges in machine learning: how to enhance the utility of datasets while preserving their privacy. Addressing this issue is critical for realizing the potential of IoT-based federated learning in healthcare and achieving robust, decentralized AI systems at the edge.
\par
Federated learning is uniquely suited to this task as it can operate on non-iid (non-independent and identically distributed) data \cite{RN139} and is an ideal match for the IoT environment since it is naturally distributed, with each participant training their own model and receiving a representative model after aggregation. This allows for each model to be trained on a smaller subset of the data, requiring less computational complexity than training a model on a complete dataset.
\par
Currently, federated learning uses either a centralized server for model aggregation or a decentralized mechanism, predominantly via blockchain \cite{RN140}. Centralized servers are less complex but require trust; this may be suitable when all models belong to the server's owner but causes issues with privacy concerns when the participating model's owners have differing goals, for example, a private hospital, a public hospital and a medical research team. Therefore, decentralized (blockchain) federated learning has the most potential. However, while applications such as cryptocurrency are enhanced by the transparency inherent in blockchain, this hinders the privacy for federated learning. Not only are all models publicly visible, and therefore can be subjected to model inversion attacks, the model suppliers may be malicious and, using current approaches, it is difficult to detect malicious models and determine whether incorrect information has been supplied alongside the model, for example federated averaging \cite{RN27} requires the participants to include the number of data samples the model has seen which is unverifiable.
\par
The fundamental concept of blockchain is the consensus mechanism, how does a collective decide what the truth is. Interestingly, this is closely related to the issue of spotting weak or even malicious networks that would corrupt the federated aggregation. While it is possible to combine consensus mechanisms with algorithms such as multi-KRUM\cite{RN116} to reduce the effect malicious models have on the final aggregated model, they do so by elimination. Unfortunately, the distinction between malicious models and non-conforming models is impossible; obviously, it is beneficial to remove malicious models but non-conforming models may have useful knowledge. Unfortunately, they may appear malicious as their dataset could be quite different from the majority, a consequence of using non-iid datasets. Additionally, these methods still require a consensus mechanism such as Proof of Work (PoW) or Proof of Stake (PoS) that are not designed with federated learning in mind.
\par
The trade-off for decentralizing federated learning by utilizing blockchain is transparency; every block in the chain is publicly available, making privacy a significant challenge. Validating a participating model requires domain-specific input, which may be difficult to obtain or unavailable for sharing. This is especially true in the medical field, where sharing raw images is unacceptable. One approach to mitigate this issue is to encode the input data; however, it is crucial that this data cannot be decoded, whether by brute force or a model inversion attack. Techniques such as homomorphic encryption excel in this regard but necessitate increased computational complexity to enable the matrix multiplication required for neural networks and is incompatible with non-linear activation functions such as rectified linear units (ReLU).
\par
Recent advances in multimodal federated learning (FL) have demonstrated that privacy-preserving techniques such as differential privacy can be integrated to handle heterogeneous data modalities across distributed nodes \cite{RN174}. In parallel, IoT healthcare research is increasingly exploring generative AI-driven human digital twins \cite{RN175} and mobile AIGC-enabled personalization \cite{RN177}, which require robust privacy and efficient on-device computation. PoR’s privacy-preserving blockchain validation complements these approaches by enabling secure, modality-agnostic aggregation without centralizing sensitive data.
\par
Furthermore, scalability in federated learning is a persistent challenge, particularly in large-scale IoT healthcare deployments with thousands of devices. Techniques such as hierarchical FL, model compression, and blockchain sharding have been proposed to mitigate communication and storage overheads, as seen in large IoT-based digital twin systems \cite{RN175, RN177}. PoR can incorporate these strategies to maintain low per-round latency and storage growth while ensuring secure and verifiable aggregation.
\par
To ensure the privacy of the data while enabling federated learning, we utilize masked autoencoders (MAE) \cite{RN127} to generate a feature map and continue to mask a high percentage of the input \cite{RN131} (unlike classical MAE, which does not mask the input outside of training), we obtain an encoded subarray of patches from the original image. This method provides high semantic content to the downstream classifier, maintains accuracy, and is highly resistant to attempts to reconstruct the original, unencoded data.
\par
This paper focuses on a privacy-enhancing, resilient paradigm, enabling efficient utilization of IoT resources to support machine learning training at the edge for healthcare/medical applications. We propose a novel framework that integrates a consensus mechanism designed solely for federated learning, named Proof of Reasoning (PoR), that utilizes masked autoencoders (MAE) \cite{RN127} to allow privacy-aware data sharing and minimize computational complexity. We summarize our contributions as follows:

\begin{itemize}
    \item We introduce PoR, a novel consensus mechanism that mitigates the effect of malicious models and is a part of the federation, as opposed to PoW, while resulting in similar guarantees of the infeasibility of replacing the entire chain with false data.
    \item We show that masked autoencoders can continue to use a high masking proportion in downstream tasks without catastrophic effect on the accuracy of linear-probing and, as a result, allow the use of the masked and encoded data for validation and aggregation of the downstream part of the model.
    \item Using both masked autoencoders and PoR, we can use more complex aggregation methods when generating the global (aggregated) model from the participating (local) models.
\end{itemize}

%% file: 02_Introduction/02_system_overview.tex
\input{figures/MAE_Overview}

The following sections delve into our system's operation in detail; however, Figure \ref{fig:MAE_Overview} provides an illustrative example of how the final federated blockchain learning system operates with three participants, each employing distinct methods for training their local networks.
\begin{enumerate}
    \item \textbf{Data Encoding:} Each participant encodes their data, potentially collected intraoperatively, using their individually trained encoder transformer (Step 1). The encoded data is then passed to their downstream classifier for immediate use (Step 2), while simultaneously serving as input for the next round of local training (Step 3).
    \item \textbf{Local Training Variations:}
        \begin{itemize}[leftmargin=0.845em]
            \item \textit{Participant 1} conducts training entirely on their local infrastructure, leveraging computational resources directly within their institution.

            \item \textit{Participant 2} utilizes a distributed IoT mist network to train their classifier. Devices in the network may employ standard distributed training algorithms, such as the All-Reduce algorithm, or engage in local federated learning. Alternatively, each IoT device can act as an independent participant, as each processes only a subset of the encoded data.

            \item \textit{Participant 3} outsources the training of their downstream classifier to another institution. They share only the encoded data and their local model, ensuring privacy while leveraging external computational resources.
        \end{itemize}
        
    \item \textbf{Blockchain Integration:} After completing local training, each participant uploads an Encoder-Decoder Interface (EDI) transaction to the blockchain (Step 4). The EDI transaction includes all necessary information as detailed in Section \ref{ssec:FedTransaction}. If Participant 2 treats each IoT device as an independent participant, multiple EDI transactions, one per device, are submitted to the blockchain.

    \item \textbf{Consensus and Validation:} The blockchain employs the Proof of Reasoning (PoR) consensus mechanism to validate each uploaded network against an encoded example datapoint provided by the participant (Step 5). This step ensures the integrity of contributions while maintaining privacy.
    
    \item \textbf{Ranking and Aggregation:} The validated networks are ranked (Step 6) based on their performance across all uploaded datapoints, following the specified aggregation policy $\pi$. The ranked networks are then aggregated into a global model (Step 7), which is distributed back to each participant, initiating the next round of local training.
\end{enumerate}

This decentralized, privacy-preserving system enables robust collaboration across multiple institutions and IoT devices, optimizing resource utilization and enhancing model performance while safeguarding sensitive patient data.

%% file: figures/MAE_Overview.tex
\begin{figure}[t]
\centering
\hspace*{-2em}
\includegraphics[width=1.1\columnwidth]
{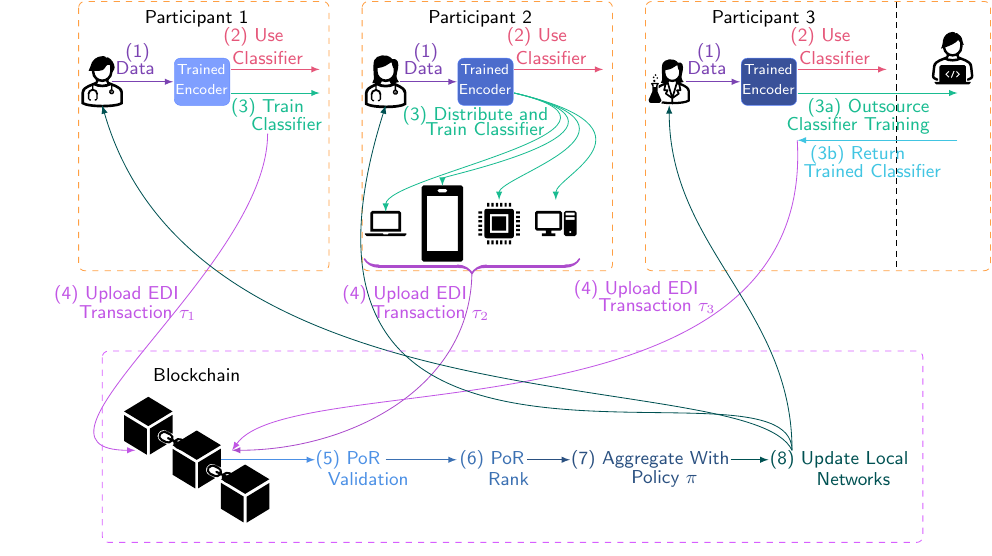}
\caption{An example of PoR being used by three participants.}
\label{fig:MAE_Overview}
\end{figure}

%% file: 03_RelatedWork/01_related_work.tex
\section{Related Work}

\subsection{Differential Privacy}
\input{03_RelatedWork/02_differential_privacy}

\subsection{Homomorphic Encryption}
\input{03_RelatedWork/03_homomorphic_encryption}

\subsection{Functional Encryption}
\input{03_RelatedWork/04_functional_encryption}

%% file: 03_RelatedWork/02_differential_privacy.tex
Differential privacy is a robust mathematical framework that protects any single datapoint in the dataset by guaranteeing that the inclusion or exclusion of said datapoint does not significantly affect the output of the computation on the dataset, thereby protecting each individual's privacy \cite{RN144, RN145}. This is achieved by introducing controlled randomness or noise to the data or, in the case of machine learning, the model's parameters, preventing an adversary from inferring sensitive information about any individual.
\par
In machine learning, differential privacy is commonly applied during model training to safeguard against privacy breaches in sensitive domains like healthcare. Techniques such as differentially private stochastic gradient descent (DP-SGD) \cite{RN143} adapt the traditional Stochastic Gradient Descent (SGD) algorithm by including two extra steps: gradient clipping and noise addition. Gradient clipping is applied to the gradient vector ($g(x_i)$) at each step of training to bind the influence of each individual datapoint by scaling the gradient if the $\ell_2$ norm is greater than $C$, the gradient norm bound. 

\begin{displaymath}
    \overline{g}(x_i)
    \left\{
    \begin{array}{l}
      \ltnorm{g(x_i)} \leq C = g(x_i)\\
      \ltnorm{g(x_i)} > C = g(x_i) / \frac{\ltnorm{g(x_i)}}{C}
    \end{array}
  \right.
\end{displaymath}

Following gradient clipping, each gradient vector has a noise signal applied before summing and aggregating the gradients into the final gradient vector for that epoch. However, clipping combined with added noise reduces model accuracy and is further exacerbated in a federated learning paradigm due to the non-iid nature of each participant's dataset. Furthermore, gradient clipping is negatively affected by the number of participants.
\par
In principle, PoR could be integrated with differential privacy at the aggregation or local training stage, for example, using DP-SGD \cite{RN143} or the differentially private federated multi-task learning framework described in \cite{RN177}, allowing adjustable privacy budgets per participant. Such integration could further mitigate privacy leakage without altering PoR’s consensus logic. In multimodal FL scenarios, this approach aligns with works applying DP to multi-source data while preserving model utility \cite{RN174}.

%% file: 03_RelatedWork/03_homomorphic_encryption.tex
Homomorphic encryption allows computations to be performed directly on encrypted data, without requiring decryption, making it highly valuable in privacy-preserving machine learning, where sensitive data can remain encrypted throughout the model training and inference processes, ensuring data confidentiality. While fully holomorphic encryption (FHE) schemes, such as CKKS \cite{RN148}, enable operations such as addition and multiplication, they introduce computational overhead, as encrypted operations are typically slower than their plaintext counterparts. Furthermore, matrix multiplication (the backbone of machine learning) requires a large computation overhead \cite{RN146} and FHE cannot handle non-linear activation functions, such as ReLU, as encrypted values cannot be compared to zero, and requires the use of Self-Learnable Activation Functions (SLAF) \cite{RN147}.

%% file: 03_RelatedWork/04_functional_encryption.tex
For a given datapoint and a function to be applied to said datapoint, functional encryption allows the owner to encrypt their data and generate a decryption key that will apply the aforementioned function to the encrypted datapoint, resulting in the same, plaintext, output that would have been achieved if the original function was applied to the plaintext datapoint \cite{RN149}. As opposed to FHE, there is no need to decrypt the output of the function; however, the same issues regarding increased complexity and issues with matrix multiplication still apply. Additionally, the owner of the data can place restrictions on the usable functions, causing issues of permission and dataset compatibility.

%% file: 04_Methods/01_methods.tex
\section{Materials and methods}

\input{figures/MAE_vs_PoR}

Proof of Reasoning (PoR) is a consensus mechanism that enables advanced, verifiable aggregation in federated learning. To provide privacy-preserving and efficient edge processing, we integrate a downsized masked autoencoder (MAE) as a feature extractor, followed by a lightweight downstream classifier. Only the encoder is deployed at the edge, reducing training and inference cost while retaining key information. For aggregation, each participant submits its classifier weights, one encoded datapoint, the classifier’s output, and the ground truth to the blockchain.
\par
Unlike standard MAEs (Figure \ref{sfig:VMAE}), our design processes only unmasked patch encodings, thereby providing a theoretical guarantee against model inversion attacks. More specifically, only a small subset of encoded patches are seen by the downstream federated models, generated by the masked autoencoder (MAE). These encodings lack sufficient spatial or semantic information to permit image reconstruction; thus, at most only the subset of encoded image patches may be reconstructed, not the original input. The decoder, used only during pre-training, adds complexity without affecting deployment.

\subsection{Pre-training}
\input{04_Methods/02_pre_training}

\input{figures/MAE_Training}

\subsection{Training}
\input{04_Methods/03_training}

\subsection{Federation}
\input{04_Methods/04_federation}

%% file: figures/MAE_vs_PoR.tex
\begin{figure}[t]
\begin{subfigure}[T]{\columnwidth}
\centering
\includegraphics[width=0.97\textwidth]
{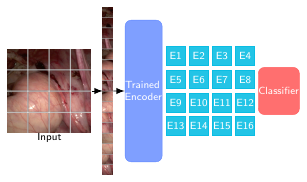} 
\caption{Vanilla MAE}
\label{sfig:VMAE}
\end{subfigure}
\begin{subfigure}[B]{\columnwidth}
\centering
\includegraphics[width=0.97\textwidth]
{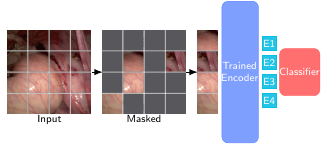}
\caption{PoR MAE}
\label{sfig:PoRMAE}
\end{subfigure}
\caption{MAE Training and upstream feature map usage. In contrast to vanilla MAE, where the input is split into unmasked patches which are then fed sequentially into a trained encoder resulting in an encoding per patch, PoR MAE continues to mask the input, as it was when training the encoder, and only the remaining (unmasked) patches are fed to the trained encoder resulting in significantly fewer encoded patches being passed to the classifier.}
\label{fig:MAE_Comparison}
\end{figure}


%% file: 04_Methods/02_pre_training.tex
Inputs are augmented via resizing and random cropping. We tested random masking and correlation-based masking; in both MAE training and downstream classification, random masking with a normal distribution gave superior results. For a masking ratio $\mu_r$ and total number of patches $\rho$, $\mu_r \cdot \rho$ patches are replaced with a learnable mask token.
\par
Both encoder and decoder use Vision Transformer (ViT) blocks, but the encoder is reduced to 26.2M parameters (~30\% of ViT-Base) by processing patches individually, while the decoder has 85.5M parameters to handle masked and unmasked patches simultaneously. Since the decoder is used only in pre-training, its size does not impact deployment. The model was trained on a 4-core CPU/4GB GPU laptop, reflecting realistic hospital resources (Table \ref{table:VitVsEncDec}).
\par
Following MAE conventions, positional embeddings are added to patch projections, with mask tokens projected for masked patches. Unmasked patches are processed by the encoder, reassembled with masked tokens, and decoded for reconstruction. Mean squared error (\lossfn{MSE}) loss on masked patches matched MAE \cite{RN127}, outperforming mean absolute error (\lossfn{MAE}\footnotemark) in downstream classification accuracy.

\begin{table}[t]
\centering
 \begin{tabular}{c c c c}
 Model Name & Heads & Width & Layers \\
 \hline
 ViT-Huge & 16 & 1280 & 32 \\
 ViT-Large & 16 & 1024 & 24 \\
 ViT-Base & 12 & 768 & 12 \\
 Edi-Encoder & 1 & 1280 & 2 \\
 Edi-Decoder & 1 & 192 & 2
 \end{tabular}
  \caption{Encoder and Decoder Transformer parameters vs standard ViT}
  \label{table:VitVsEncDec}
 \end{table}

\footnotetext{We use \lossfn{X} for error functions in order to differentiate Masked Autoencoders (MAE) from Mean Absolute Error (\lossfn{MAE})}

%% file: figures/MAE_Training.tex
\begin{figure}[t]
\centering
\includegraphics[width=\columnwidth]
{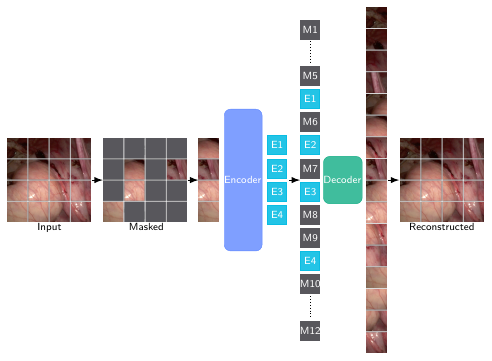} 
\caption{Self-supervised training process of a masked autoencoder (MAE): The input image is divided into patches, with a high percentage masked out. The unmasked patches are assigned positional encodings and processed by the encoder, which transforms them into encoded representations. These encoded patches are then combined with masked patches, each containing a learnable mask token and its positional encoding, and passed to the decoder, which reconstructs the original image. This training enables the encoder to extract meaningful representations from limited input data.}
\label{fig:MAETrain}
\end{figure}

%% file: 04_Methods/03_training.tex
\subsubsection*{Downstream Classifier Design}
The downstream classifier is a compact feed-forward network with batch normalization, global average pooling, and a final dense layer. Unlike conventional MAE pipelines, masking is maintained at inference, reducing input size, computational cost, and attack surface. Encoded and masked datapoints uploaded for validation are obfuscated and resistant to model inversion, yet remain useful for ranking and aggregation in PoR.

%% file: 04_Methods/04_federation.tex
\input{figures/Blockchain_MAE}
\subsubsection*{Federated Transaction}
\label{ssec:FedTransaction}
For each participant $i \in \mathcal{P}$, Proof of Reasoning (PoR) introduces an Encoder-Decoder Interface (EDI) transaction $\tau_i$ that is uploaded to the blockchain (Figure \ref{fig:MAE_Blockchain}). Each EDI transaction encapsulates the weights of the participant’s downstream classifier, $\omega_i$, and a list of one or more tuples, where each tuple contains an encoded datapoint $\kappa_i$, the output of the downstream classifier for that datapoint $\hat{y}_i$, and the actual classification label $y_i$. Let $\Omega_i$ denote a downstream classifier with weights $\omega_i$ then: 
\begin{align*}
    \tau_i &= (\omega_i, [(\kappa_i, \hat{y}_i, y_i),\cdots]); & 
    \hat{y}_i &= \Omega_i (\kappa_i); & 
    \kappa_i &= \lambda_i(x_i)
\end{align*}
Where $\lambda_i$ represents a masked encoder and $x_i$ is the original input. The EDI transaction does not include the masked encoder $\lambda_i$ or the raw input $x_i$. Even with access to the transaction data $(\omega_i, [\kappa_i, \hat{y_i}, y_i), \cdots])$, it is computationally infeasible to derive $x_i$ by inverting $\lambda$ due to the one-way nature of the encoding process. 
\par
This structure enhances federated learning by enabling more sophisticated and verifiable aggregation methods. Unlike federated averaging (FedAvg) \cite{RN27}, which relies on unverifiable counts of training examples used by each participant, the EDI transaction provides transparency and robustness, paving the way for novel aggregation schemes.
\par
Per-round storage grows O(NR) with participants N and rounds R; however, pruning older rounds or off-chain storage with on-chain hashes can mitigate this. PoR validation scales linearly with the number of participants and, for a large N, batching or hierarchical consensus may be applied.

\subsubsection*{Proof of Reasoning}
The PoR mechanism begins by verifying that each network in the participant pool $\mathcal{P}$ produces outputs within a specified tolerance $\epsilon$, ensuring a baseline level of trustworthiness. While this step does not entirely eliminate the risk of malicious networks, since inputs could originate from a different domain, allowing potentially malicious networks to pass validation, the subsequent aggregation step is designed to minimize their impact. Each network is then scored and ranked according to an aggregation policy $\pi$, which operates on the set of all EDI transactions, $\mathcal{T}$, recorded on the blockchain. The aggregation policy $\pi$ comprises two primary functions: a scoring function $\mathcal{S}^\pi: \mathbb{R}^n \times \mathbb{R} \rightarrow \mathbb{R}$, which evaluates the performance of each classifier $\Omega_i$ on all of the included encoded datapoints $\bigcup_{j \in \mathcal{P}}\{\kappa_j\}$, and a reduction function $\mathcal{R}^\pi: \mathbb{R}^n \rightarrow \mathbb{R}$, which consolidates these scores into a final ranking for each participant. Specifically, for each participant i:

\begin{align}
    \text{PoR}_{\text{Validation}} & = |\Omega_i(\kappa_i) - \hat{y}_i| < \epsilon \\
    \text{PoR}_{\text{Rank}} & = \mathcal{R}^\pi(\{\mathcal{S}^\pi(\Omega_i(\kappa_j), y_j) | j \in \mathcal{P}\})
\end{align}
A straightforward example of an aggregation policy for multiclass classification is a linear scoring method based on classification probabilities. In this approach, each network is evaluated by assigning points proportional to the placement of its predicted probability for the correct class. For instance, if a network's predicted probability for the correct class is the third highest among all predictions, it is awarded 
(num\_classes - 2) points. The reduction function then aggregates these scores by summing them across all inputs to generate an overall ranking for each network.
\par
PoR operates orthogonally to existing consensus methods (potentially complementing Aggregation Schemes such as multi-KRUM) offering transparency and accountability by directly linking each model’s contribution to its measured accuracy on shared encoded inputs, rather than unverifiable self-reported statistics as in FedAvg. This output-based scoring enables fairer, more robust aggregation strategies, addressing key shortcomings of conventional methods and improving reliability in privacy-sensitive domains such as healthcare. Unlike approaches such as FedAvg with PoS or FedSGD with PoW, PoR embeds verifiable validation into the consensus process, avoiding reliance on untrusted metadata. While PoW/PoS add latency without improving model quality verification \cite{RN116}, PoR simultaneously validates and ranks models using encoded samples, reducing both consensus delays and the need for external trust.
\par
Beyond resisting model inversion, PoR’s ranking and cross-validation reduce the risk of poisoning or replacement attacks by testing the performance of each model and discarding or assigning a low rank to malicious or degraded models, negating their effect on the aggregated model. This can be enhanced via anomaly detection or signature checks. Its decentralised design supports continuous learning as data or regulations change, allowing participants to join or leave without disrupting global updates.

%% file: figures/Blockchain_MAE.tex
\begin{figure}[t]
\centering
\includegraphics[width=\columnwidth]
{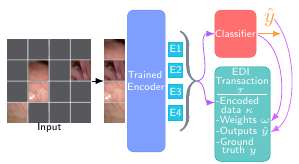}
\caption{An example of generating an Encoder-Decoder Interface (EDI) transaction to be added to the blockchain using PoR. The unmasked patches of a single datapoint are encoded by the participant's trained encoder transformer and added to the transaction as the encoded data array $\kappa$. Additionally, the weights of the downstream classifier $\omega$, the output of the downstream classifier on the encoded data $\hat{y}$ and the true classification of the input $y$ are also added to the transaction.}
\label{fig:MAE_Blockchain}
\end{figure}

%% file: 05_Results/01_results.tex
\section{Results and discussion}
\label{sec:results}

\subsection{Cifar10 Experiments}
\input{05_Results/02_Cifar10_Experiments}

\subsection{Transfer Learning Experiments}
\input{05_Results/03_Transfer_Learning_Experiments}

\subsection{Chest and Pneumonia Mnist Experiments}
\input{05_Results/04_Chest_Pneumonia_Mnist}

%% file: 05_Results/02_Cifar10_Experiments.tex
\input{Tables/EncoderDecoderParams}

As this framework is designed with the IoT infrastructure of a hospital in mind the first requirement was an encoder-decoder architecture that could run on a leaner environment, namely a laptop with decent computational power but not to the same extent as a dedicated machine learning server. This limited the number of heads and layers that our Encoder-Decoder transformers could use. Table \ref{table:EncDecParams} shows the effect on the reconstruction accuracy of various transformer parameters on the Cifar-10 dataset \cite{RN113}.

\input{Tables/EDI_MaskingLossAndMaskMethod}
\input{Tables/EDI_Reconstruct_results}
Since our use of the masked autoencoder (MAE) differs from the original design \cite{RN127}, after selecting optimal transformer settings, we explored combinations of training parameters (i.e. masking percentage, masking method, and loss function) in order to evaluate their effect on the downstream classifier. We found that higher masking percentages reduced reconstruction accuracy but improved downstream classification, with 90\% masking giving the best trade-off. Using \lossfn{MAE} as the loss function yielded slightly higher reconstruction accuracy but significantly reduced downstream performance. Correlation-based masking (selecting the most similar, most dissimilar, or a 50/50 mix of patches) consistently degraded results, whether used for training the encoder or for feature map generation. On the other hand, random masking, despite producing non-deterministic outputs, proved superior results; this can be attributed to the stochastic regularization effect of random masking, forcing the encoder to generalize from diverse and unpredictable patch subsets. In contrast, correlation-based masking introduces systematic biases, allowing the model to overfit to predictable missing regions. Random masking also better simulates realistic input variability, improving downstream robustness. We report the first (non-cherry-picked) results for each parameter set in Table \ref{table:Cifar10MaskingExperiments} and the reconstruction accuracy of the masked autoencoder using the highlighted settings on various datasets in Table \ref{table:EDI_reconst_res}.

\input{Tables/NonFederatedCifar10}

\input{Tables/SimpleFederatedCifar10}

For the downstream classifier, we started with a simple feedforward network consisting of batch normalization and global average pooling, followed by one hidden layer and one output layer. This worked well in the case of traditional (non-federated) machine learning and is viable in a paradigm where encoded datasets can be traded freely (Table \ref{table:SimpleModel_Non-Federated}); however, in a federated paradigm, the lack of data (when cifar-10 is split between two or four participants) and the variance of each datapoint is detrimental to the accuracy of the classifier, despite no drastic change in the reconstruction accuracy of the masked autoencoder and including a high dropout rate of 70\% which improved the results (Table \ref{table:SimpleModel_Federated}). However, we found that if all participants used the same masked autoencoder trained on the whole dataset, we obtained results equal to the non-federated case; this is impractical, so transfer learning was the next logical step.

%% file: Tables/EncoderDecoderParams.tex
\begin{table}[t]
\centering
 \begin{tabular}{| c c c | c c c | c |}
  \hline
 \multicolumn{3}{|c|}{Encoder} & \multicolumn{3}{c|}{Decoder} & \multirow{2}{4em}
 {Accuracy} \\
 Heads & Width & Layers & Heads & Width & Layers & \\ 
 \hline
 1     & 1024  & 2      & 1      & 192  & 2      & 15.95\% \\
 1     & 1024  & 2      & 1      & 208  & 2      & 15.99\% \\
 2     & 1024  & 4      & 2      & 128  & 4      & 15.62\% \\
 \rowcolor{lightgray} 1 & 1280 & 2 & 1 & 192 & 2 & 16.01\% \\
 2     & 1280  & 2      & 1      & 128  & 2      & 14.80\% \\
 1     & 2048  & 2      & 1      & 128  & 2      & 15.30\% \\
 \hline
 \end{tabular}
 \caption{Impact of varying the number of heads, width, and layers for both the encoder and decoder transformers: The encoder, being the core component used throughout the system, is designed to be larger than the decoder while maintaining efficiency to operate on resource-constrained IoT hardware. Both components must remain lightweight to ensure compatibility with edge devices. Each configuration was evaluated using a masking ratio of 75\% on images from the Cifar-10 dataset.}
  \label{table:EncDecParams}
 \end{table}

%% file: Tables/EDI_MaskingLossAndMaskMethod.tex
 \begin{table}[t]
\centering
 \begin{tabular}{c c c c c}
    Mask \% & Masking Method & Loss Function & Loss & Accuracy \\
 \hline
 60\% & Random & \lossfn{MSE} & 0.0103 & 18.14\% \\
 60\% & Random & \textbf{\lossfn{MAE}} & \textbf{0.0680} & \textbf{18.45\%} \\
 62.5\% & Random & \lossfn{MSE} & 0.0110 & 17.72\% \\
 75\% & Random & \lossfn{MSE} & 0.0142 & 16.08\% \\
 85\% & Random & \lossfn{MSE} & 0.0191 & 13.83\% \\
 85\% & Correlation & \lossfn{MSE} & 0.0259 & 9.87\% \\
  \rowcolor{lightgray} 90\% & Random & \lossfn{MSE} & 0.0233 & 12.35\% \\
 95\% & Random & \lossfn{MSE} & 0.0307 & 10.32\%
 \end{tabular}
  \caption{Effect of masking percentage, masking method, and loss function on the reconstruction accuracy of the masked autoencoder (MAE): The experiments were conducted after training for 100 epochs on the Cifar-10 dataset (training set size: 40,000, testing set size: 10,000). While using \lossfn{MAE} led to higher reconstruction accuracy, employing \lossfn{MSE} with a high masking percentage of 90\% resulted in superior classification accuracy for the downstream classifier. This indicates that, despite reduced reconstruction performance, the encoder learned a more effective representation for classification of the input data.}
  \label{table:Cifar10MaskingExperiments}
 \end{table}

%% file: Tables/EDI_Reconstruct_results.tex
\begin{table}[t]
\centering
 \begin{tabular}{c c c c c}
 Dataset & No. Examples & MSE & MAE & Accuracy \\
 \hline
 Cifar-10 & 40,000 & 0.0233 & 0.1101 & 12.35\% \\
 Cifar-100 & 40,000 & 0.0238 & 0.1110 & 13.32\% \\
 ImageNet & 60,000 & 0.0281 & 0.1245 & 8.63\% \\
 ChestMnist & 80,000 & 0.0063 & 0.0525 & 51.58\%
 \end{tabular}
  \caption{Results after 100 epochs on training a Masked Autoencoder on various datasets with a mask\% of 90\%}
  \label{table:EDI_reconst_res}
 \end{table}

%% file: Tables/NonFederatedCifar10.tex
\begin{table}[t]
    \begin{subtable}[t]
    {\columnwidth}
    \centering
    \begin{tabular}{| m{1.15em} c | m{2.45em} | m{3.25em} |} 
    \hline
    \multicolumn{2}{|c|}{Hidden} & \multirow{2}{2.5em}{Mask\%} & \multirow{2}{3.5em}{Accuracy} \\
    Layers & Width & & \\
    \hline
         1 & 2560      & 50\% & 62.91\% \\
         1 & 4096      & 75\% & 59.90\% \\
         \rowcolor{lightgray} 1 & 5120 & 50\% & 64.06\% \\
         1 & 5120      & 62.5\% & 61.00\% \\
         1 & 5120      & 75\% & 60.32\% \\
         1 & 8192      & 75\% & 59.24\% \\
         2 & $\displaystyle \binom{\text{5120}}{\text{256}}$& 75\% & 58.73\% \\

         \hline
    \end{tabular}
     \caption{Effect of hidden layer configurations and input masking percentage on downstream classifier accuracy: The simplified downstream classifier (excluding the residual bottleneck layer) was evaluated with varying numbers of hidden layers, layer widths, and input masking percentages. Each configuration was trained on 40,000 images from the Cifar-10 dataset using stochastic gradient descent (SGD) for network optimisation.}
        \label{table:Full_Data_SimpleModel_Non-Federated}
     \end{subtable}
     \begin{subtable}[b]
     {\columnwidth}
    \centering
    \begin{tabular}{| m{3em} | m{3.25em} | m{2.5em} | m{2.5em} |} 
        \hline
        Range & Accuracy & Top 2 & Top 3 \\
        \hline
        0-25   & 49.05\% & 70.55\% & 81.88\% \\
        25-50  & 50.26\% & 71.13\% & 82.35\% \\
        50-75  & 50.13\% & 71.40\% & 81.91\% \\
        75-100 & 50.04\% & 70.37\% & 81.29\% \\
        \hline
    \end{tabular}
     \caption{Classification accuracy for the simplified downstream classifier using a masking percentage of 62.5\%. Each classifier was trained on 10,000 images from the Cifar-10 dataset (40,000 images split between 4 models) using the ADAM optimiser.}
        \label{table:Split_Model_Non-Federated}
     \end{subtable}
  \caption{Non-Federated Cifar-10 Downstream results.}
  \label{table:SimpleModel_Non-Federated}
 \end{table}

%% file: Tables/SimpleFederatedCifar10.tex
 \begin{table}[t]
\centering
 \begin{tabular}{c c c c c}
  Range & \lossfn{MSE} & Accuracy & Top 2 & Top 3 \\
 \hline
 0-25   & 1.3641 & 51.65\% & 72.42\% & 83.47\% \\
 25-50  & 1.3518 & 51.51\% & 72.95\% & 83.45\% \\
 50-75  & 1.3617 & 51.50\% & 72.13\% & 82.86\% \\
 75-100 & 1.3545 & 51.72\% & 72.27\% & 82.86\% \\  
 \end{tabular}
  \caption{Results after 5 federation rounds of 25 epochs each, with 25 pre-federated epochs on Cifar-10. We include a dropout rate of 70\% after the single hidden layer.}
  \label{table:SimpleModel_Federated}
 \end{table}

%% file: 05_Results/03_Transfer_Learning_Experiments.tex
\input{Tables/Inet_Transfer_Cifar10}

\input{figures/ResBNL}

After training each masked autoencoder on a different subset of 60,000 images from the imagenet dataset \cite{RN152} for 100 epochs, we proceeded to continue training each model on a subset of Cifar-10 for a further 100 epochs before training the downstream model on the same subset of Cifar-10 (Table \ref{table:InetTransferCifar10}).

\input{Tables/Inet_Transfer_FederatedCifar10}

To improve the federated results, a combination of changes were required. First, we applied transfer learning by freezing all the masked autoencoder's weights, apart from the encoder transformer. Next, we added a single (full pre-activation \cite{RN31}) residual bottleneck layer (Figure \ref{fig:ResBNL}), which downsamples the input by $\frac{1}{8}$ before restoring the input shape, in front of the global averaging layer (the batch normalization layer is a part of the residual bottleneck layer and is therefore redundant) and include dropout of 70\% after the hidden layer. Finally, we aggregated the encoder transformer of each participant before the first aggregation step; this preserves the obfuscating ability of the masked autoencoder while improving the performance considerably despite a slight increase in complexity (Table \ref{table:InetTransfer_FederatedCifar10}).

%% file: Tables/Inet_Transfer_Cifar10.tex
\begin{table}[t]
\centering
 \begin{tabular}{|c c c c c c|}
   \hline
  Range  & epochs & Transfer Type & Accuracy & Top 2 & Top 3 \\
  \hline
  0-100  &  100   &     None      &  60.99\% & 79.69\% & 88.14\% \\
  0-100  &  100   &  Encoder Only &  61.57\% & 79.42\% & 88.07\% \\
  0-100  &  100   &     Full      &  61.02\% & 78.99\% & 87.56\% \\
  \hline
  0-25   &   25   &  Encoder Only &  50.72\% & 72.02\% & 82.93\% \\
  25-50  &   25   &  Encoder Only &  50.81\% & 71.28\% & 82.45\% \\
  50-75  &   25   &  Encoder Only &  49.57\% & 70.61\% & 81.51\% \\
  75-100 &   25   &  Encoder Only &  49.80\% & 71.04\% & 82.18\% \\
  \hline
  0-50   &   25   &  Encoder Only &  54.14\% & 74.55\% & 84.84\% \\
  50-100 &   25   &  Encoder Only &  54.34\% & 75.08\% & 84.37\% \\
  \hline
 \end{tabular}
  \caption{Impact of transfer learning strategies on downstream classifier accuracy for Cifar-10: The masked autoencoder (MAE) was pre-trained on ImageNet, and transfer learning was applied by freezing all weights except for the encoder transformer (Encoder Only) before further training on Cifar-10. This approach outperformed transfer learning without freezing any weights (Full), with both methods yielding superior results compared to training without transfer learning (None).}
  \label{table:InetTransferCifar10}
 \end{table}

%% file: figures/ResBNL.tex
\begin{figure}[t]
\centering
\includegraphics[width=0.95\columnwidth]
{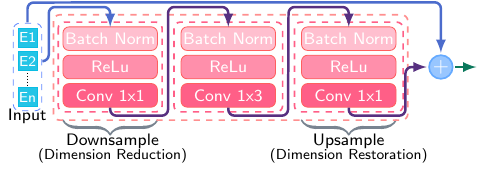} 
\caption{The structure of the residual bottleneck layer: The layer consists of three sequential sub-layers. First, a 1D convolution is applied to reduce the input size, with the number of features reduced to $\frac{1}{8}$ of the original. Next, the second sub-layer performs 1D convolution with a kernel size of 3, maintaining the reduced feature dimensions. Finally, the third sub-layer restores the feature dimensions to their original size using a 1D convolution, restoring the number of features to its original size. The output of this sequence is then added to the original input, creating a residual connection to enhance feature learning.}
\label{fig:ResBNL}
\end{figure}

%% file: Tables/Inet_Transfer_FederatedCifar10.tex
\begin{table}[t]
\centering
 \begin{tabular}{|c c c c c|}
 \hline
  Range & \lossfn{MSE} & Accuracy & Top 2 & Top 3 \\
 \hline
 0-25   & 1.2852 & 56.36\% & 75.73\% & 85.73\% \\
 25-50  & 1.3150 & 54.85\% & 74.93\% & 84.56\% \\
 50-75  & 1.3349 & 55.31\% & 74.82\% & 84.57\% \\
 75-100 & 1.2862 & 56.18\% & 76.03\% & 85.81\% \\
 \hline
 0-50   & 1.1971 & 60.43\% & 79.15\% & 87.63\% \\  
 50-100 & 1.2444 & 60.03\% & 78.76\% & 87.35\% \\
 \hline
 \end{tabular}
  \caption{Cifar-10 classification accuracy (comparable to our previous results in \cite{RN132} using FedAvg + PoW). Each participant's downstream classifier is trained for 25 epochs and aggregated. Thereafter, each participant is trained for 5 epochs before aggregation for 25 rounds total.}
  \label{table:InetTransfer_FederatedCifar10}
 \end{table}

%% file: 05_Results/04_Chest_Pneumonia_Mnist.tex
While Cifar-10 is a popular benchmark dataset for Computer Vision, medical images are very different semantically and visually. Furthermore, unlike standard digital images, medical scans such as CAT and MRI require processing to convert from their units into a format appropriate for viewing digitally. For example, CAT scans are measured in Hounsfield units (HU) with values between -1,024 and 3,071 representing the absorption of radiation, and require windowing and leveling to restrict the values into the standard 8-bit grayscale range of 0-255. 
\par
\input{Tables/Pneumonia_FederatedVsNon}

In contrast to multi-channel images, single-channel images did not suffer from the same issues, nor did they require the encoder transformer to be federated at any point, retaining the maximal amount of privacy and resistance to attacks. After training each masked autoencoder on a different subset of the ChestMnist dataset \cite{RN157}, requiring no transfer learning (the two image domains are close to identical), we then trained the downstream binary classification task using the relatively small (5,000 training images) PneumoniaMnist dataset \cite{RN150} without any federation of the masked autoencoders ensuring that each participant has a distinct encoder transformer. By using a low epoch count in between aggregation rounds of 5 epochs and a larger number of rounds (15 in total) the federated results outperformed the non-federated case, by at least 2.56\% (Table \ref{table:Pneumonia_FederatedVSNon}). Our Pneumonia classifier uses a dropout of 15\% after the hidden layer but is otherwise identical to the classifier used in the Cifar-10 and imagenet experiments. The same range is used for both datasets, i.e. the whole dataset for ranges 0-100 and 25\% for the other ranges, with the federated models being trained for 25 epochs prior to the first aggregation step.

%% file: Tables/Pneumonia_FederatedVsNon.tex
\begin{table}[t]
    \centering
    \begin{tabular}{| m{2.85em} m{2.35em} m{2.35em} c c c c |} 
    \hline
    Range & Epochs & Rounds & \lossfn{MSE} & Accuracy & Recall & Precision \\
    \hline
    0-100  & 100 & N/A & 1.1354 & 83.81\% & 96.67\% & 81.08\% \\
    0-25   & 100 & N/A & 0.8677 & 83.81\% & 97.44\% & 80.68\% \\
    25-50  & 100 & N/A & 1.6054 & 80.61\% & 98.97\% & 76.74\% \\
    50-75  & 100 & N/A & 1.4703 & 83.49\% & 98.97\% & 79.59\% \\
    75-100 & 100 & N/A & 0.6741 & 87.02\% & 93.33\% & 86.87\% \\
    \hline
    0-100  &  25 & N/A & 0.9001 & 82.05\% & 97.69\% & 78.72\% \\
    0-25   &  25 & N/A & 0.7228 & 81.73\% & 97.95\% & 78.28\% \\
    25-50  &  25 & N/A & 0.7187 & 81.89\% & 97.18\% & 78.79\% \\
    50-75  &  25 & N/A & 0.9720 & 80.61\% & 98.72\% & 76.85\% \\
    75-100 &  25 & N/A & 0.7121 & 82.05\% & 96.67\% & 79.20\% \\
    \hline
    \multicolumn{7}{|c|}{Federated} \\
    \hline
    0-25   &  5 & 15 & 0.5253 & 84.62\% & 87.95\% & 87.50\% \\
    25-50  &  5 & 15 & 0.8450 & 81.57\% & 95.38\% & 79.32\% \\
    \rowcolor{lightgray} 50-75  &  5 & 15 & 0.5370 & 89.58\% & 94.62\% & 89.35\% \\
    75-100 &  5 & 15 & 0.5533 & 88.78\% & 95.13\% & 87.91\% \\
    \hline
    \end{tabular}
     \caption{Binary classification accuracy of the downstream classifier on the PneumoniaMNIST dataset: Each participant utilized a unique masked autoencoder (MAE) trained independently on distinct subsets of the ChestMNIST dataset. The encoder transformers were not federated, ensuring unique representations for each participant. Transfer learning was not required, highlighting the effectiveness of domain-specific training.}
  \label{table:Pneumonia_FederatedVSNon}
 \end{table}

%% file: 06_Conclusion/01_conclusion.tex
\section{Conclusion}
The transformer architecture has long been a staple in natural language processing (NLP) but has only recently gained prominence in vision tasks, with Vision Transformers (ViTs) demonstrating exceptional performance in image classification \cite{RN128}. Self-supervised learning has emerged as an invaluable tool for pre-training, particularly in the medical field, where labelling data often requires highly skilled experts or is subject to significant delays, as is the case with conditions like anastomotic leaks. These challenges constrain the rate of labelled data collection. By leveraging the self-supervised capabilities of masked autoencoders (MAEs) and continuing to mask inputs when training downstream classifiers, this system enables the secure sharing of datapoints with enhanced privacy, facilitates federated learning-specific blockchain validation methods, and mitigates the risks associated with model inversion attacks.
\par
In contrast to the research in \cite{RN134}, where federated networks outperformed the non-federated equivalent, multi-channel datasets pose significant challenges due to their increased data requirements. Fully federating the masked autoencoder could alleviate these data limitations but comes at the cost of reduced privacy and diminished resistance to model inversion attacks. A balanced solution was identified by aggregating the weights of each participant's encoder transformer before initiating federated learning. This compromise maintains a balance between privacy, attack resistance, and performance. However, as single-channel images do not face these issues, further investigation is warranted; for example, training one MAE per channel and concatenating their output encoding, masking channels instead of patches or applying preprocessing techniques are all suitable research directions for further study.
For instance, when Vision Transformers (ViTs) are applied directly for classification, a learnable classification token is prepended to the set of image patches \cite{RN128}, enabling the model to aggregate global contextual information. While this approach allows the encoder to learn task-specific features, it necessitates the use of a labelled dataset during pre-training, potentially limiting its applicability in scenarios with scarce annotated data.
\par
For greyscale images, MAEs have shown promising results when trained in a self-supervised manner on domain-specific, unlabelled datasets, subsequently serving as feature detectors for downstream classifiers trained on smaller labelled datasets. This approach is particularly suited for scenarios demanding high privacy and fidelity, where labelled data is scarce but unlabelled data is more accessible within the same domain.
\par
MAEs also have potential for image generation tasks \cite{RN121}, enabling dataset augmentation in scenarios where domain-specific data is limited. However, image generation introduces new privacy challenges, as the transparency of model weights and architecture on the blockchain exposes the system to potential attacks. Addressing these challenges will require novel approaches to privacy enhancement.

%% file: IoTJ_2024.bib
@inproceedings{RN143,
author = {Abadi, Martin and others},
title = {Deep learning with differential privacy},
booktitle = {Proceedings of the 2016 ACM SIGSAC conference on computer and communications security},
pages = {308-318},
year = {2016},
type = {Conference Proceedings}
}

@techreport{RN113,
author = {Alex, Krizhevsky and Geoffrey, Hinton},
title = {CIFAR-10 (Canadian Institute for Advanced Research)},
institution = {Toronto University},
url = {https://www.cs.toronto.edu/~kriz/learning-features-2009-TR.pdf},
year = {2009},
type = {Report}
}

@inproceedings{RN149,
author = {Boneh, Dan and Sahai, Amit and Waters, Brent},
title = {Functional Encryption: Definitions and Challenges},
booktitle = {TCC 2011},
series = {Theory of Cryptography},
publisher = {Springer Berlin Heidelberg},
pages = {253-273},
ISBN = {978-3-642-19571-6},
year = {2011},
type = {Conference Proceedings}
}

@article{RN140,
author = {Brecko, Alexander and Kajati, Erik and Koziorek, Jiri and Zolotova, Iveta},
title = {Federated Learning for Edge Computing: A Survey},
journal = {Applied Sciences},
volume = {12},
number = {18},
pages = {9124},
ISSN = {2076-3417},
url = {https://www.mdpi.com/2076-3417/12/18/9124},
year = {2022},
type = {Journal Article}
}

@article{RN132,
author = {Calo, James and Lo, Benny},
title = {Federated Blockchain Learning at the Edge},
journal = {Information},
volume = {14},
number = {6},
pages = {318},
ISSN = {2078-2489},
url = {https://www.mdpi.com/2078-2489/14/6/318},
year = {2023},
type = {Journal Article}
}

@inproceedings{RN134,
author = {Calo, J. and Lo, B.},
title = {IoT Federated Blockchain Learning at the Edge},
booktitle = {IEEE EMBC},
pages = {1-4},
ISBN = {2694-0604},
DOI = {10.1109/EMBC40787.2023.10339946},
year = {2023},
type = {Conference Proceedings}
}

@article{RN93,
author = {Chen, J. and Ran, X.},
title = {Deep Learning With Edge Computing: A Review},
journal = {Proceedings of the IEEE},
volume = {107},
number = {8},
pages = {1655-1674},
ISSN = {1558-2256},
DOI = {10.1109/JPROC.2019.2921977},
year = {2019},
type = {Journal Article}
}

@article{RN175,
author = {Chen, J. and Shi, Y. and Yi, C. and Du, H. and Kang, J. and Niyato, D.},
title = {Generative-AI-Driven Human Digital Twin in IoT Healthcare: A Comprehensive Survey},
journal = {IEEE Internet of Things Journal},
volume = {11},
number = {21},
pages = {34749-34773},
ISSN = {2327-4662},
DOI = {10.1109/JIOT.2024.3421918},
year = {2024},
type = {Journal Article}
}

@article{RN177,
author = {Chen, Jiayuan and others},
title = {A revolution of personalized healthcare: Enabling human digital twin with mobile AIGC},
journal = {IEEE network},
volume = {38},
number = {6},
pages = {234-242},
ISSN = {0890-8044},
year = {2024},
type = {Journal Article}
}

@inproceedings{RN148,
author = {Cheon, Jung Hee and Kim, Andrey and Kim, Miran and Song, Yongsoo},
title = {Homomorphic Encryption for Arithmetic of Approximate Numbers},
booktitle = {ASIACRYPT 2017},
series = {Advances in Cryptology – ASIACRYPT 2017},
publisher = {Springer International Publishing},
pages = {409-437},
ISBN = {978-3-319-70694-8},
year = {2017},
type = {Conference Proceedings}
}

@article{RN152,
author = {Chrabaszcz, Patryk and Loshchilov, Ilya and Hutter, Frank},
title = {A downsampled variant of imagenet as an alternative to the cifar datasets},
journal = {arXiv preprint arXiv:1707.08819},
year = {2017},
type = {Journal Article}
}

@inproceedings{RN130,
author = {Deng, J. and Dong, W. and Socher, R. and Li, L. J. and Kai, Li and Li, Fei-Fei},
title = {ImageNet: A large-scale hierarchical image database},
booktitle = {2009 IEEE Conference CVPR},
pages = {248-255},
ISBN = {1063-6919},
DOI = {10.1109/CVPR.2009.5206848},
year = {2009},
type = {Conference Proceedings}
}

@article{RN131,
author = {Devlin, Jacob and Chang, Ming-Wei and Lee, Kenton and Toutanova, Kristina},
title = {Bert: Pre-training of deep bidirectional transformers for language understanding},
journal = {arXiv preprint arXiv:1810.04805},
year = {2018},
type = {Journal Article}
}

@article{RN128,
author = {Dosovitskiy, Alexey and others},
title = {An image is worth 16x16 words: Transformers for image recognition at scale},
journal = {arXiv preprint arXiv:2010.11929},
year = {2020},
type = {Journal Article}
}

@inproceedings{RN145,
author = {Dwork, Cynthia},
title = {Differential Privacy},
booktitle = {ICALP 2006},
series = {Automata, Languages and Programming},
publisher = {Springer Berlin Heidelberg},
pages = {1-12},
ISBN = {978-3-540-35908-1},
year = {2006},
type = {Conference Proceedings}
}

@inproceedings{RN144,
author = {Dwork, Cynthia and McSherry, Frank and Nissim, Kobbi and Smith, Adam},
title = {Calibrating Noise to Sensitivity in Private Data Analysis},
booktitle = {TCC 2006},
series = {Theory of Cryptography},
publisher = {Springer Berlin Heidelberg},
pages = {265-284},
ISBN = {978-3-540-32732-5},
year = {2006},
type = {Conference Proceedings}
}

@article{RN174,
author = {Feng, C. and Feng, D. and Huang, G. and Liu, Z. and Wang, Z. and Xia, X. G.},
title = {Robust Privacy-Preserving Recommendation Systems Driven by Multimodal Federated Learning},
journal = {IEEE TNNLS},
volume = {36},
number = {5},
pages = {8896-8910},
ISSN = {2162-2388},
DOI = {10.1109/TNNLS.2024.3411402},
year = {2025},
type = {Journal Article}
}

@inproceedings{RN127,
author = {He, Kaiming and Chen, Xinlei and Xie, Saining and Li, Yanghao and Dollár, Piotr and Girshick, Ross},
title = {Masked autoencoders are scalable vision learners},
booktitle = {Proceedings of the IEEE/CVF conference CVPR},
pages = {16000-16009},
year = {2022},
type = {Conference Proceedings}
}

@inproceedings{RN138,
author = {He, Kaiming and Zhang, Xiangyu and Ren, Shaoqing and Sun, Jian},
title = {Deep residual learning for image recognition},
booktitle = {Proceedings of the IEEE conference on computer vision and pattern recognition},
pages = {770-778},
year = {2016},
type = {Conference Proceedings}
}

@inproceedings{RN31,
author = {He, Kaiming and Zhang, Xiangyu and Ren, Shaoqing and Sun, Jian},
title = {Identity Mappings in Deep Residual Networks},
booktitle = {Computer Vision – ECCV 2016},
series = {Computer Vision – ECCV 2016},
publisher = {Springer International Publishing},
pages = {630-645},
ISBN = {978-3-319-46493-0},
year = {2016},
type = {Conference Proceedings}
}

@article{RN150,
author = {Kermany, Daniel S. and others},
title = {Identifying Medical Diagnoses and Treatable Diseases by Image-Based Deep Learning},
journal = {Cell},
volume = {172},
number = {5},
pages = {1122-1131.e9},
ISSN = {0092-8674},
DOI = {https://doi.org/10.1016/j.cell.2018.02.010},
url = {https://www.sciencedirect.com/science/article/pii/S0092867418301545},
year = {2018},
type = {Journal Article}
}

@article{RN139,
author = {Konečný, Jakub and McMahan, Brendan and Ramage, Daniel},
title = {Federated optimization: Distributed optimization beyond the datacenter},
journal = {arXiv preprint arXiv:1511.03575},
year = {2015},
type = {Journal Article}
}

@article{RN136,
author = {Krizhevsky, Alex and Sutskever, Ilya and Hinton, Geoffrey E.},
title = {ImageNet classification with deep convolutional neural networks},
journal = {Commun. ACM},
volume = {60},
number = {6},
pages = {84–90},
ISSN = {0001-0782},
DOI = {10.1145/3065386},
url = {https://doi.org/10.1145/3065386},
year = {2017},
type = {Journal Article}
}

@inproceedings{RN121,
author = {Li, Tianhong and Chang, Huiwen and Mishra, Shlok and Zhang, Han and Katabi, Dina and Krishnan, Dilip},
title = {Mage: Masked generative encoder to unify representation learning and image synthesis},
booktitle = {Proceedings of the IEEE/CVF Conference CVPR},
pages = {2142-2152},
year = {2023},
type = {Conference Proceedings}
}

@misc{RN27,
author = {McMahan, H. Brendan and Moore, Eider and Ramage, Daniel and Hampson, Seth and Arcas, Blaise Agüera y},
title = {Communication-Efficient Learning of Deep Networks from Decentralized Data},
journal = {Proceedings of the 20 th International Conference on Artificial Intelligence and Statistics (AISTATS) 2017. JMLR: W\&CP volume 54},
volume = {54},
month = {20--22 April},
year = {2017},
type = {Conference Paper}
}

@misc{RN146,
author = {Rizomiliotis, Panagiotis and Triakosia, Aikaterini},
title = {On Matrix Multiplication with Homomorphic Encryption},
journal = {Proceedings of the 2022 on Cloud Computing Security Workshop},
pages = {53–61},
publisher = {Association for Computing Machinery},
DOI= {10.1145/3560810.3564267},
url = {https://doi.org/10.1145/3560810.3564267},
year = {2022},
type = {Conference Paper}
}

@inproceedings{RN58,
author = {Ronneberger, Olaf and Fischer, Philipp and Brox, Thomas},
title = {U-net: Convolutional networks for biomedical image segmentation},
booktitle = {MICCAI},
publisher = {Springer},
pages = {234-241},
year = {2015},
type = {Conference Proceedings}
}

@article{RN135,
author = {Tanghatari, Ehsan and Kamal, Mehdi and Afzali-Kusha, Ali and Pedram, Massoud},
title = {Distributing DNN training over IoT edge devices based on transfer learning},
journal = {Neurocomputing},
volume = {467},
pages = {56-65},
ISSN = {0925-2312},
DOI = {https://doi.org/10.1016/j.neucom.2021.09.045},
url = {https://www.sciencedirect.com/science/article/pii/S0925231221014235},
year = {2022},
type = {Journal Article}
}

@inproceedings{RN157,
author = {Wang, Xiaosong and Peng, Yifan and Lu, Le and Lu, Zhiyong and Bagheri, Mohammadhadi and Summers, Ronald M},
title = {Chestx-ray8: Hospital-scale chest x-ray database and benchmarks on weakly-supervised classification and localization of common thorax diseases},
booktitle = {Proceedings of the IEEE conference CVPR},
pages = {2097-2106},
year = {2017},
type = {Conference Proceedings}
}

@article{RN147,
author = {Xiong, Jichao and Chen, Jiageng and Lin, Junyu and Jiao, Dian and Liu, Hui},
title = {Enhancing privacy-preserving machine learning with self-learnable activation functions in fully homomorphic encryption},
journal = {JISA},
volume = {86},
pages = {103887},
ISSN = {2214-2126},
DOI = {https://doi.org/10.1016/j.jisa.2024.103887},
url = {https://www.sciencedirect.com/science/article/pii/S2214212624001893},
year = {2024},
type = {Journal Article}
}

@article{RN116,
author = {Yang, Zhanpeng and Shi, Yuanming and Zhou, Yong and Wang, Zixin and Yang, Kai},
title = {Trustworthy federated learning via blockchain},
journal = {IEEE Internet of Things Journal},
volume = {10},
number = {1},
pages = {92-109},
ISSN = {2327-4662},
year = {2022},
type = {Journal Article}
}
